\documentclass[11pt,a4paper]{emulateapj}
\usepackage{natbib}
\usepackage{amssymb, amsmath, amsbsy}%, rotating}
\usepackage{color}
\usepackage{rotating}

\usepackage{epsfig}
\usepackage{amsmath}
\usepackage{amssymb}
\usepackage{natbib}
\usepackage{subfigure}
\usepackage{graphicx}
\slugcomment{Accepted by {\it The Astrophysical Journal Letter}}

\newcommand{\civ}{C {\sc iv}}

\newcommand{\oii}{O {\sc ii}}

\newcommand{\zabs}{z_{\rm{abs}}}

\begin{document}

\def\mean#1{\left< #1 \right>}

\title{Probing the Metal Enrichment of the Intergalactic Medium at $z=5-6$ Using the Hubble Space Telescope}
\author{Zheng Cai\altaffilmark{1,8}, Xiaohui Fan\altaffilmark{2}, Romeel Dave\altaffilmark{3,4,5}, Kristian Finlator\altaffilmark{6}, Ben Oppenheimer\altaffilmark{7}}
\affil{$^1$ UCO/Lick Observatory, University of California, 1156 High Street, Santa Cruz, CA 95064, USA;\  zcai@ucolick.org }
\affil{$^2$ Steward Observatory, University of Arizona, 933 North Cherry Avenue, Tucson, AZ, 85721, USA}
\affil{$^3$ Institute for Astronomy, University of Edinburgh, Royal Observatory, Edinburgh, EH9 3HJ}
\affil{$^4$ University of the Western Cape, Bellville, Cape Town 7535, South Africa}
\affil{$^5$ South African Astronomical Observatories, Observatory, Cape Town 7925, South Africa}
\affil{$^6$ New Mexico State University, Las Cruces, NM 88003, USA}
\affil{$^7$ CASA, Department of Astrophysical and Planetary Sciences, University of Colorado, 389 UCB, Boulder, CO 80309, USA}
\affil{$^{8}$ Hubble Fellow}

%\affil{$^8$ National Optical Astronomical Observatory, Tucson, AZ, 85719, USA}

\altaffiltext{1} {Email: zcai@ucolick.org}

\begin{abstract}
{ We test the galactic outflow model by probing associated galaxies of four strong intergalactic CIV absorbers at $z=5$--6 using the Hubble Space Telescope (HST) ACS ramp narrowband filters. The four strong CIV absorbers reside at $z=5.74$, $5.52$, $4.95$, and $4.87$, with column densities ranging from $N_{\rm{CIV}}=10^{13.8}$ cm$^{-2}$ to $10^{14.8}$ cm$^{-2}$. At $z=5.74$, we detect
an i-dropout Ly$\alpha$ emitter (LAE) candidate with a projected impact parameter of 42 physical kpc from the CIV absorber.
This LAE candidate has a Ly$\alpha$-based star formation rate (SFR$_{\rm{Ly\alpha}}$) of 2 $M_\odot$ yr$^{-1}$ and a UV-based SFR of  4 $M_\odot$ yr$^{-1}$. {Although we cannot completely rule out that this $i$-dropout emitter may be an [OII] interloper, its measured properties are consistent with the CIV powering galaxy at $z=5.74$. } For CIV absorbers at $z=4.95$ and $z=4.87$, although we detect two LAE candidates with impact parameters of 160 kpc and 200 kpc, such distances are larger than that predicted from the simulations. Therefore we treat them as non-detections. 
For the system at $z=5.52$, we do not detect LAE candidates, placing a 3-$\sigma$ upper limit of SFR$_{\rm{Ly\alpha}}\approx 1.5\ M_\odot$ yr$^{-1}$.  In summary, in these four cases, we only detect one plausible CIV source at $z=5.74$. Combining the modest SFR of the one detection and the three non-detections, 
our HST observations strongly support that smaller galaxies (SFR$_{\rm{Ly\alpha}} \lesssim 2\ M_\odot$ yr$^{-1}$) 
are main sources of intergalactic CIV absorbers, and such small galaxies play a major role in the metal enrichment of the intergalactic medium at $z\gtrsim5$. }
\end{abstract}

\section{Introduction}

The detection of metal absorption systems at $z > 6$ indicates that the intergalactic medium (IGM) has already been enriched by the end of reinization \cite[e.g.,][]{becker06, becker12, keating14}. 
A  key question is to understand what the sources of early metals are and how they got distributed throughout the IGM in the post-reionization universe.   
Theoretical models suggest that galactic superwinds from star-forming galaxies is a main source of the early IGM enrichment. 
\citet{oppenheimer06} explored high-redshift IGM absorbers in cosmological simulations, employing various prescriptions for 
galactic outflows to enrich the IGM at $z = 2 - 6$. 
\citet{oppenheimer09} and \citet{finlator15} studied the relation between galaxies and CIV absorbers in simulations, and 
predicted that strong absorbers are the result of outflows from galaxies with stellar masses $M= 10^{7-9}$ M$_\odot$, 
and star formation rates (SFRs) of a few $M_\odot$ yr$^{-1}$. 
The separation between the CIV absorber and the associated galaxy is tens of kpc ($\approx5-50$ kpc) from the quasar line of sight. 
{ Note the virial radius of such small galaxies is $\approx 7-8$ kpc at $z\approx5-6$ (e.g., Wechsler et al. 2002). }

At $z >5$, observations of IGM - galaxy interactions are difficult.  In the field of the strong CIV absorber J1030+0524 at $z=5.7$,  
\citet{diaz11} reached a minimum UV-based star formation of $\sim 5\ M_\odot$ yr$^{-1}$ with deep VLT observations. They identified a galaxy associated with CIV absorber 
at an impact parameter of 79 kpc. \citet{diaz14, diaz15} further present the large-scale overdensity of Ly$\alpha$ emitters (LAEs) at $z\approx5.7$ associated with this CIV absorption 
systems. This survey reached a Ly$\alpha$-based star formation rate (SFR$_{\rm{Ly\alpha}}$) of 5 $M_\odot$ yr$^{-1}$ and the closest CIV-galaxy pair had an impact parameter 
of 300 (213 $h^{-1}$) kpc. Nevertheless, these great efforts have not ruled out the possibility that fainter galaxies with small impact parameters may be the source of CIV absorption. 
To better test theoretical simulations, we need to probe galaxies down to stellar masses of $M\lesssim10^{9}$ M$_\odot$ and SFRs of $\approx 1-2 \ M_\odot$ yr$^{-1}$.

Using HST/ACS narrowband ramp filters,  we carried out deep narrow-band imaging of four IGM CIV absorbers at $z =5 -6$. 
We adjusted the ACS ramp filters to the desired wavelength for Ly$\alpha$ detection to search for LAEs in the vicinity of CIV absorbers. 
The ACS ramp filter has a monochromatic patch of $40"\times20"$, corresponding to an area of 250 kpc $\times$ 125 kpc (physical). 
With the ACS ramp filters, we are able to reach $3\sigma$ SFR$_{\rm{Ly\alpha}}$ of $=1-2\ M_\odot$ yr$^{-1}$. 
{ Typical ($M^*$) Lyman break galaxies (LBGs) at $z\approx5-6$ have SFRs of $\sim 7\ M_\odot$ yr$^{-1}$. The depth probed in these observations is much lower than 
the typical LBGs and previous ground-based observations. This depth also allows us to test the theoretical models of galactic outflows (e.g., Oppenheimer et al. 2009) 
which predict that the CIV sources have SFRs of $\lesssim1$ M$_\odot$ yr$^{-1}$. }
% This paper is structured as follows. 
In \S2, we introduce the HST observations. In \S3, we discuss our observational results and the target selection of galaxy candidates. In \S4, we discuss the 
implication of the observations.   We convert redshifts to physical distances assuming a $\rm{\Lambda}$CDM cosmology with $\Omega_m= 0.3$, 
$\Omega_{\Lambda}=0.7$ and $h=0.70$ ($h_{70}$).  Throughout this paper, we use physical distances and AB magnitude.

\section{Hubble Space Telescope Observations}

\subsection{Sample Selection}

The CIV absorbers at $z=5-6$ were selected from the sample of \citet{simcoe11}. From this sample, we select four secure, strong CIV absorbers at $z=4.87 - 5.74$ in two QSO sight lines: SDSSJ1030+0524 ($z_{\rm{QSO }} = 6.30$) and SDSSJ1306+0356 ($z_{\rm{QSO }}= 6.00$). The CIV column densities range from $13.8 $ cm$^{-2}$ to $14.8$ cm$^{-2}$. These two QSO fields have broadband (F775W, F850LP) images available from \citet{stiavelli05}. 
The sample is summarized in Table 1.

\subsection{Observations}

%Using HST ramp filter listed in Table~1, we observe each CIV absorber, with the central wavelength tuned to the corresponding Ly$\alpha$ emission line in order to search for Ly$\alpha$ emission from galaxies associated with CIV absorption lines. 
We devoted 15 orbits to observe four CIV absorber fields in different filters (HST-GO-12860). We used five orbits using FR853N ramp filter for the $z=5.7$ system, four orbits using F782N for the $z=5.5$ absorber, and three orbits using F716N for each of the two $z=4.9$ systems. { For each ramp filter, the central wavelength is tuned to the CIV absorption to search for the Ly$\alpha$ emission associated with CIV absorption. The ramp filter has a peak efficiency (70\%) spanning 3000 km s$^{-1}$, much larger than the systematic velocity offset of Ly$\alpha$ to CIV absorber (450 km s$^{-1}$, Steidel et al. 2010). }
Our HST observations allow us to measure: (1) Ly$\alpha$-based star formation rate (SFR$_{\rm{Ly\alpha}}$), (2) projected distances between the galaxies and CIV absorbers, and (3) UV-based star formation rates (SFR$_{\rm{UV}}$) or their upper limits. 
The observations probe SFR$_{\rm{Ly\alpha}}=$ 2 $M_\odot$ yr$^{-1}$ at the 4-$\sigma$ level, corresponding to a Ly$\alpha$ flux of $2.2 \times 10^{42}$ erg s$^{-1}$  \citep{kennicutt98}. %(Kennicutt et al. 1998). 
 Data reduction is performed with Multidrizzle \citep{koekemoer13}. %,  
Using a uniform 0.6" diameter aperture, we calculated the depth of our observations in different bands. The narrowband magnitude for point sources reached the $5\sigma$ level of $m_{\rm{F716N}}=26.1$,  $m_{\rm{F782N}}= 26.0$, and $m_{\rm{F853N}}=26.1$.
The broadband observations reached $m_{\rm{F850LP}}= 27.8$ and $m_{\rm{775W}}=27.7$ at the 5-$\sigma$ level. 
 Photometry is performed with SExtractor \citep{bertin96} (Bertin \& Arnouts 1996) using the root-mean-square map converted from the inverse 
variance image generated by Multidrizzle \citep{casertano00}.  %and $m_{\rm{F775W}}= 28.3$ in 3-$\sigma$ %(AB magnitude in this paper.)

\section{Results}

%\subsection{Galaxy template model}

\subsection{Selection of Ly$\alpha$ emitter candidates}

We select Ly$\alpha$ emitter (LAE) candidates with rest-frame equivalent width (EW) $>20$ \AA. The EW of 20 \AA\ corresponds to the color selection 
criteria of $m_{\rm{broadband}} - m_{\rm{narrowband}} > 0.65$. %At $z>5.7$, we expect 
Also, we require the signal-to-noise ratio (SNR) of the continuum-subtracted flux to be greater than 3. The signal-to-noise is calculated using $m_{\rm{B}}- 
m_{\rm{NB}} \ge \Sigma \times \sqrt{\sigma^2_{\rm{NB}}+ \sigma^2_{\rm{B}}}$, where $\Sigma=3.0$, and $m_{\rm{NB}}$, $\sigma_{\rm{NB}}$, $m_{\rm{B}}$, 
$\sigma_B$ representing the  magnitudes and errors of the narrowband and broadband.  The selection criteria are shown in Fig.~1, Fig.~2, and Fig.~3. Galaxies 
reside above both red and blue dotted lines are selected.  One LAE candidate satisfies the selection criteria in each of the three CIV absorber fields 
 at $z=5.74$, $z=4.95$, and $z=4.87$. We visually checked individual exposures of these candidates to make sure: (a) their photometry 
are not affected by cosmic rays; (b) sources do not reside at the detector edges where the photometry is less accurate. We do not detect any LAE candidates associated 
with the CIV system at $z=5.52$. 
%We summarize %We summarize all the LAE candidates in Table~1. 

In each figure, we use the following galaxy model to fit the photometry of LAE candidates to derive the flux of Ly$\alpha$ emission, the Ly$\alpha$-based and UV-based SFR. 
%We construct the galaxy continuum model using HyperZ (Bolzonella et al. 2000). The HyperZ program fits the \citet{bruzual03} stellar synthesis 
%models to the spectral energy distribution (SED) of galaxies, including broadband photometry of WFC3 F775W and F850LP bands. 
 We use the \citet{bruzual03} models with a Salpeter Initial Mass Function, a 0.2 solar metallicity ($Z = 0.004$), a continuous star formation history, and an age of 30 Myr \cite[e.g.,][]{ono10}. We use the Ly$\alpha$ optical depth derived by \citep{madau95} to account for the IGM absorption. % blueward the Ly$\alpha$ emission. 
 Then, we add the dust extinction following \citet{calzetti00} with $R_V = A(V)/E(B - V)=3.1$ %(e.g., Oyarzun et al. 2010) 
 for the $z\approx5.7$ candidates, and set $R_V$ a free parameter for two $z\approx5.0$ candidates. 
 Further, we added the Ly$\alpha$ emission based on our narrowband observations,  assuming the FWHM of Ly$\alpha$ is $300$ km s$^{-1}$ (e.g., Jiang et al. 2013).

\subsection{Physical Properties of LAE candidates}

In Table~1, we provide the physical properties of the three LAE candidates. %In this section, we briefly summarize their properties. 
For the CIV absorber at $\zabs= 5.744$, we detect a LAE candidate with an impact parameter of 42 kpc. %The  coordinate of the LAE is $\alpha$: 10:30:26.746,  $\delta$: +05:24:59.76. 
%This LAE candidate has a 3-$\sigma$ upper limit of F775W magnitude of $m_{\rm{775W}} > 28.30$,  F850LP magnitude of $m_{\rm{850LP}}= 27.10 \pm0.13$, and FR853N magnitude is $m_{\rm{853N}}= 25.90\pm0.21$.  %The,. 
This LAE candidate has an expected Ly$\alpha$ luminosity of $L_{\rm{Ly\alpha}}= 1.75 \pm 0.44 \times 10^{42}$ erg s$^{-1}$, corresponding to the SFR$_{\rm{Ly\alpha}}$ of $2.0$ M$_\odot$ yr$^{-1}$ \cite[e.g.,][]{kennicutt98}. %Since we have the UV continuum measurement redward of Ly$\alpha$ emission, we can also measure the UV-based SFR (SFR$_{\rm{UV}}$), and 
%The 
The SFR$_{\rm{UV}}$ is estimated to be 4.1 M$_\odot$ yr$^{-1}$, using the flux density of the best-fit SED around $\lambda_{\rm{rest}}=1500$ \AA\ \citep{madau98}. 
Diaz et al. (2011) found a LAE at 79 kpc away from the CIV absorber and suggested that this LAE could be the powering source.  We do not detect any NB-excess of this source in our deeper HST observations. This source has $m_{\rm{F853N}} >26.30$, $m_{\rm{F775W}}=26.75\pm0.09$, $m_{\rm{F850LP}} = 26.23\pm0.06$, which is inconsistent with a LAE at $z\approx5.7$. %in detecting Ly$\alpha$ emission than the spectroscopy presented in Diaz et al. (2011). 

For the CIV absorber at $\zabs= 4.948$, the impact parameter of the LAE candidate is 163 kpc. 
The Ly$\alpha$ luminosity is $L_{\rm{Ly\alpha}} = 3.27\pm0.56\times10^{42}$ erg s$^{-1}$, corresponding to a SFR$_{\rm{Ly\alpha}}=3.0\ M_\odot$ yr$^{-1}$. 
The SFR$_{\rm{UV}}$ is 8.7 $M_\odot$ yr$^{-1}$. For the CIV absorber at $\zabs= 4.866$, there is a LAE candidate with an impact parameter of 205 kpc. 
The Ly$\alpha$ luminosity is $L_{\rm{Ly\alpha}} = 2.72\pm0.55\times 10^{42}$ erg s$^{-1}$, corresponding to a SFR$_{\rm{Ly\alpha}}=2.5\ M_\odot$ yr$^{-1}$. The UV-based SFR is estimated to be 2.1 $M_\odot$ yr$^{-1}$. If we further assume a stellar mass ($M_*$) -- SFR relation at $z=4-5$ \cite[e.g.,][]{sparre14},  the SFR$_{\rm{UV}}$-derived stellar mass of these three LAE candidates have stellar masses of $\sim 10^{9.0-9.5}$ $M_\odot$ and halo masses of $\sim 10^{10}$ $M_\odot$. %We summarize the physical properties of each LAE candidate in Table~1. 

For the CIV absorber at $\zabs= 5.517$, we do not detect any LAE associated with this absorber, 
placing a 3$\sigma$ upper limit on the Ly$\alpha$ luminosity of $1.60\times10^{42}$ erg s$^{-1}$ and a 3-$\sigma$ upper limit on the SFR$_{\rm{Ly\alpha}}\approx 1.5\ M_\odot$ yr$^{-1}$.

The main contaminants of high-$z$ LAEs are strong [\oii] emitters at lower redshift. For two \civ\ absorbers at $z\approx 4.9$, we currently have no imaging observations bluer than the Ly$\alpha$ emission.  We expect that the contamination 
rate due to low-$z$ [\oii] emission is relatively high. [\oii] emitters at $z=0.91$ ($z=0.94$) with rest-frame EW $>61.3$ \AA\  are contaminants of LAEs at $z=4.948$ ($z=4.866$) with rest-frame EW $>20$ \AA.  From the [\oii] luminosity function at $z\approx 1.0$ \citep{takahashi07, zhu09}, strong [\oii] emitters that can be detected in our depth (EW $>61.3$ \AA, $L_{\rm{[OII]}}>10^{41.0}$ erg s$^{-1}$)  have a number density of 0.009 Mpc$^{-3}$, corresponding to $\approx0.2$ per field.  For LAE candidates at  $z=5.7$, a number of studies have been conducted at similar redshifts \cite[e.g.,][]{shimasaku06, ouchi08, ouchi17}. Spectroscopic confirmations suggest that the contamination rate is $\lesssim 20$\% at our depth. Future spectroscopic follow-ups must be conducted to confirm or rule out the LAE candidates conclusively. 

\section{Discussions}

{ We find three LAE candidates in the field of 
three strong CIV absorbers at $z=5.744$, $z=4.948$, and $z=4.866$, respectively. At $z=5.74$, we detect
an i-dropout LAE candidate at an impact parameter of 42 physical kpc from the quasar sight line.
This is a plausible CIV source which has a SFR$_{\rm{Ly\alpha}}$ of 2 $M_\odot$ yr$^{-1}$ and SFR$_{\rm{UV}}=$ 4.1 $M_\odot$ yr$^{-1}$. 
At  $z=4.95$ and $z=4.87$, we detect two galaxy candidates with impact parameters of 160 kpc and 200 kpc. { Here, let us assume galaxies 
start an outflow at the average redshift when reionization occurs ($z\approx9$ from Planck Oservations, Adam et al. 2016), then at $z\approx5.7$, 
the momentum driven winds ($v\approx200$ km s$^{-1}$) travel about 90 kpc, less than the impact parameter we observed. Also, simulations (e.g., Oppenheimer et al. 2009) 
predict that impact parameters should be $\lesssim50$ kpc to yield strong CIV absorption with $N_{\rm{CIV}}\gtrsim10^{14}$ cm$^{-2}$.} %However, from both simulations and  %Thus these two LAE candidates are unlikely to be physically associated systems, placing an 3-$\sigma$ upper limit of SFR$_{\rm{Ly\alpha}}\le 1.8\ M_\odot$ yr$^{-1}$. 
We do not detect LAEs associated with the CIV absorber at $z=5.52$. We place 3-$\sigma$ upper limits of SFR$_{\rm{Ly\alpha}}\approx 1.5\ M_\odot$ yr$^{-1}$ 
in each of the three non-detection fields.} %Thus, in these four trials, our observations detect one plausible CIV associated galaxy at 42 kpc with the SFR slightly on the high end of expectations. }

\subsection{Probing \civ\ enrichment using LAEs}

One potential issue is that not all star forming galaxies have Ly$\alpha$ emission that can be selected using the LAE selection technique. \citet{schenker12} showed that 60\% galaxies with absolute magnitude of $-20.3 < M_{\rm{UV}}< -18.7$ have Ly$\alpha$ EW greater than 20 \AA. Our observations probe galaxies down to $M_{\rm{UV}}\approx -18.5$ in 3-$\sigma$,  indicating a high completeness of using LAE technique to select galaxies in our survey.   Another issue is whether LAE candidates we detected are physically associated with the CIV absorbers. Observationally, we can evaluate this issue by calculating the number of galaxies expected in random fields: if $\ll 1$ galaxies is expected within an impact parameter in random fields, then galaxies detected within these impact parameters are more likely to be sources of CIV systems \cite[e.g.,][]{diaz14}. This is because CIV absorbers are expected to be clustered with the physically-associated galaxies. 
The LAE candidate we detected at $z=5.74\pm0.07$ is the first source reported at $z>5$ to be projected $\approx 40$ kpc away from a CIV absorber. 
From the luminosity function at $z=5.7$ \citep{ouchi08, konno17}, the $L^*_{\rm{Ly\alpha}}= 6.8\times10^{42}$ erg s$^{-1}$ and $\phi^* = 7.7\times10^{-4}$ Mpc$^{-3}$. Our limiting Ly$\alpha$ luminosity is  0.26 $\times L^*_{\rm{Ly\alpha}}$. The width of the FR853N ramp filter corresponds to a redshift interval of $\Delta z = 0.14$. The expected number of finding a LAE in random fields with $L_{\rm{Lya}}\ge 0.26 \times L^*_{\rm{Ly\alpha}}$  and at impact parameters of $\le 42$ kpc is 0.01. %We can also investigate the expected number of galaxies with $L= 0.01 - 0.26\times L^*_{\rm{Ly\alpha}}$ within a radius of 42 kpc. 
The number of faint galaxies with $L= 0.01- 0.26\times L^*_{\rm{Ly\alpha}}$ (below our detection limit) and an impact parameter within 42 kpc is 0.2. Thus, there is a relatively small probability to find such a LAE in random fields. %{ Further, both SFR and impact parameters are consistent with the simulation predictions of CIV sources. }
If this LAE candidate is confirmed by future deep spectroscopy, then one can conclude that this galaxy is highly likely to be the source of the CIV absorber at $z=5.744$. %The UV-based SFR is estimated to be $4.1$ M$_\odot$ yr$^{-1}$. 

For two CIV absorbers at $z\approx5$,  the expected number of LAEs with $L_{\rm{Ly\alpha}}\ge 0.26 \times L^*_{Ly\alpha}$  and an impact parameter of $\approx200$ kpc is 0.5. 
The number of faint LAEs ($L_{\rm{Ly\alpha}}= 0.01 - 0.26 \times L^*_{\rm{Ly\alpha}}$) is expected to be $\approx 10$ in random fields.   
{ As a result, it is unlikely that the LAE candidates we detected at $z\approx 5.0$ are sources of the CIV absorbers, and thus we treat these two sight lines as non-detections}.

\subsection{Constraining IGM enrichment models}

{ From our one detection and three non-detections, we conclude that 
 strong CIV systems at $z\approx 5-6$ do not
generally lie close to galaxies identified as LAEs with
SFR$\ga 2\ M_\odot$ yr$^{-1}$. } This is surprising because strongly
star-forming galaxies are expected to be driving the strongest
outflows at these epochs, enriching the early IGM.  Models generically
predict early, widespread enrichment of
metals~\citep[e.g.][]{madau01,oppenheimer08}, such that $z\ga 5$
represents the first epoch of IGM enrichment.  Given this, the most
straightforward interpretation of our result is that early enrichment
is dominated, at least in a volumetric sense, by small galaxies
with SFR$\la 2 M_\odot$ yr$^{-1}$.  

There are two important caveats to this result.  First, our
observations cannot select LBGs without
strong Ly$\alpha$ emission.  However, as noted in the previous section, 
 LBGs and LAEs are often coincident at these epochs 
(60\% of LBGs at $z=5-6$ are LAEs) \cite[e.g.,][]{schenker12}. %(e.g., Schenker et al. 2012). 
 If the
40\% that are not LAEs are randomly associated with LBGs, then the
probability of all four CIV systems being undetected is $0.4^4$ ($<3$\%).  Second, our selection disfavors the detection of
massive dust-obscured galaxies. Such galaxies are metal-rich and
have high SFR, so potentially we could be missing these sources.
However, typical high-SFR objects at these epochs should not be heavily dust-obscured~\cite[e.g.][]{ono10,finkelstein12}.
Future sub-millimeter observations would be required to rule out this explanation conclusively.

Setting aside the caveats, our conclusion represents an important
constraint on models of early metal production and dissemination.
Such models began by post-processing metal distributions onto
cosmological simulations~\citep[e.g.][]{dave98,aguirre01}, but
the introduction of outflows into simulations by \citet{springel03}
enabled self-consistent IGM enrichment.  \citet{oppenheimer06}
showed that a momentum-driven outflow model in which small galaxies
have higher mass loading (i.e. outflow rate relative to SFR) yielded
better agreement with CIV enrichment at $z\sim 2-4$, as compared
to models where the mass loading is independent of galaxy mass.
\citet{oppenheimer09} extended these results to $z\sim 6$, and
showed that the high mass loading factors in small galaxies resulted
in early enrichment in accord with CIV data, despite a fairly
small metal volume filling factor of a couple percent.  They further
predicted that strong absorbers ($N_{\rm{CIV}}= 10^{13.8-14.8}$
cm$^{-2}$) arise from outflows from galaxies with stellar masses
$\sim10^{7-9}$ $M_\odot$ and impact parameters from a few kpc up
to 50 kpc.  %Such simulated galaxies have SFR of about $\sim$ 1--10
%$M_\odot$ yr$^{-1}$~\citep{dave06}. 
 These predictions are consistent with our observational results.

Using cosmological simulations, \citet{finlator13} introduced on-the-fly radiative transfer into
cosmological simulations and 
examined early metal enrichment in
conjuction with an inhomogeneous ionising background to
determine the impact of local UVB fluctuations on the ionisation state.  
They argued that the mass scale of halos hosting OI absorbers was 
$10-100\times$ smaller than typical LBGs at $z\sim 6$, which is consistent 
with our results if OI and \civ\ absorption arise around similar systems.
In contrast, \citet{keating16} examined early IGM enrichment in the
Sherwood and Illustris simulations, and found that galaxies associated with strong
(EW $>1$ \AA) CIV absorbers have halo masses of $M>10^{11}$ $M_\odot$,
and impact parameters of 10-30 kpc.  These results seem to be
less in agreement with our results at face value, although a more
careful comparison is warranted.  This is interesting because
Illustris utilizes a mass loading factor that increase to lower
masses even more rapidly than in momentum-driven wind models.  Hence
other factors such as local density and ionisation may be playing
a key role.  %We note that neither of these works used direct radiative
%transfer to assess the ionisation state, although \citet{keating16}
%did examine the consequences of ionizing bubbles around galaxies.
%If, for example, faint galaxies have systematically higher ionizing 
%escape fractions~\citep{alv12,wis14}, then their tendency to eject 
%preferentially more metals \emph{and} more UV light than LBGs could 
%lead them to dominate the CIV budget, consistent with our results.  
It could be that the~\citet{keating16} 
and~\citet{nelson15} models are predicting strong CIV from dust-obscured 
galaxies that would be missed by our LBG/LAE survey.

{ To summarize, we present the detection of one plausible CIV 
source at $z=5.7$ with a projected distance of 42 kpc and a SFR$_{\rm{Ly\alpha}}=1.75\pm0.44$ M$_\odot$ yr$^{-1}$. 
Also, we found two LAE candidates in the field of two CIV absorbers at $z=4.95$ and $z=4.87$ with projected 
distances of 163 and 205 kpc, respectively. But the impact parameters are much larger than that predicted 
by simulations, making them unlikely to be CIV sources. We do not detect CIV-associated LAEs at $z=5.52$. 
From our observations, 
no CIV-associated LAEs have impact parameters $\lesssim 40$ kpc with SFR$_{\rm{Ly\alpha}}\ge1.5\ M_\odot$ yr$^{-1}$
(3-$\sigma$ limit)}.  Particularly, for strongest absorbers with
EW $>$ 2\AA, we do not detect LAEs with $3\sigma$ SFR$_{\rm{Ly\alpha}}\ge1.4$
$M_\odot$ yr$^{-1}$. Our
observations provide strong constraints on models of early enrichment
in simulations.  Future facilities, including James
Webb Space Telescope ({\it JWST}), will probe IGM-galaxy relation
down to a much fainter galaxy population where some models predict
that the CIV arises.

{{\bf Acknowledgement: } { We thank the referee for valuable comments. ZC thanks Marcel Neeleman for his 
useful comments.}  ZC, XF thank the support from the US NSF grant AST 11-07682. Support for part of this work was also provided by
NASA through the Hubble Fellowship grant HST-HF2-51370 awarded by the Space Telescope Science Institute,
which is operated by the Association of Universities for 
Research in Astronomy, Inc., for NASA, under contract
NAS 5-26555. }

%\newpage

\figurenum{1}
\begin{figure}[tbp]
\epsscale{1.2}
\label{fig:02+04}
\plotone{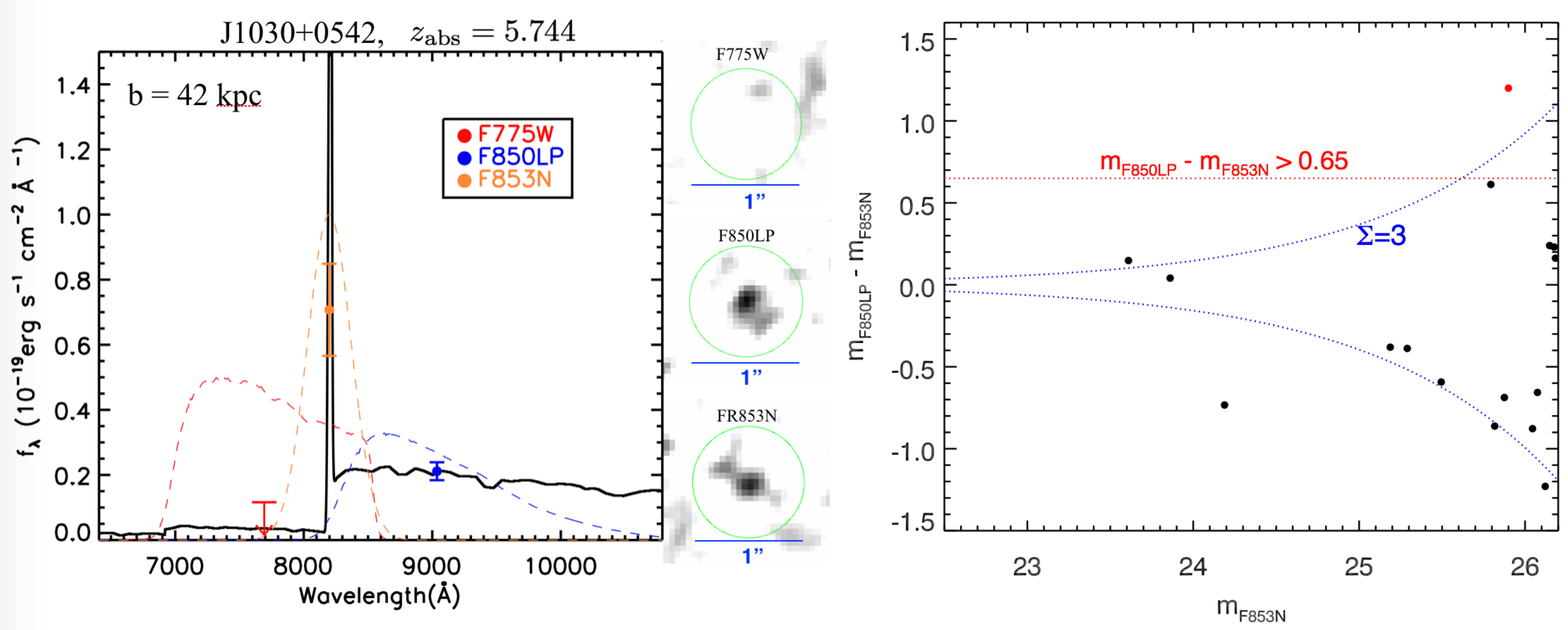}
\caption{The left panel shows the best-fit spectrum (black line) of the candidate of the Ly$\alpha$ emitter (LAE) associated with the CIV absorber at $z=5.744$. The impact parameter between this LAE candidate and the CIV absorber is 42 kpc. The filter response curves of F775W (red dashed line), F850LP (blue dashed line) and ACS ramp filter (yellow dashed line) are overplotted. In addition, the photometry in three different bands are overplotted at the effective wavelength of each filter. The right panel shows the color - magnitude diagram of galaxies within the monochromatic field of view of FR853N field. The color selection criteria for the LAE candidates (red dot) are broadband (B) - narrowband (NB) $>0,65$ (red horizontal dotted line) and $m_{\rm{B}} - m_{\rm{NB}} \ge \Sigma \times \sqrt{\sigma^2_{\rm{NB}}+ \sigma^2_{\rm{B}}}$, where $\Sigma= 3.0$ (blue dotted line).  }
\end{figure}

\figurenum{2}
\begin{figure}[tbp]
\epsscale{1.2}
\label{fig:02+04}
\plotone{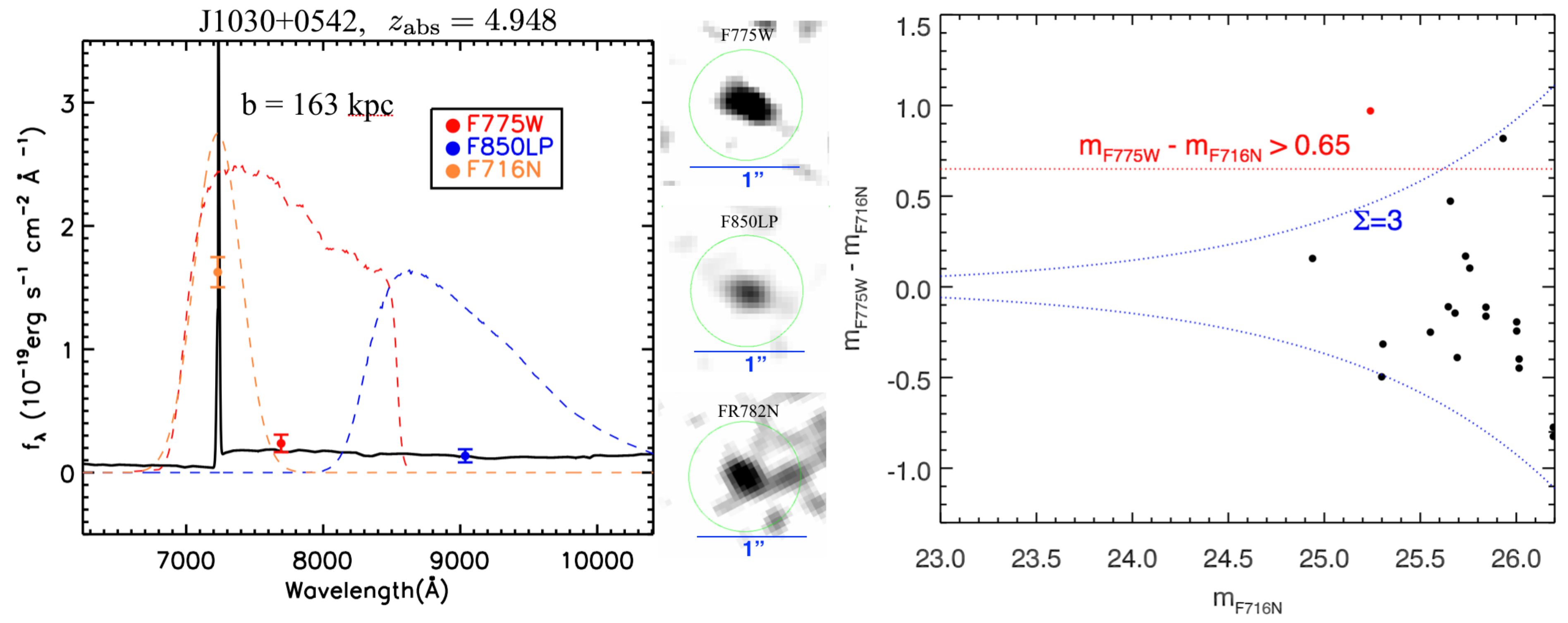}
\caption{Same format as the Figure~1, presenting the photometry and the selection of LAE candidates in the CIV absorber field at $z=4.948$. }
\end{figure}

\figurenum{3}
\begin{figure}[tbp]
\epsscale{1.2}
\label{fig:02+04}
\plotone{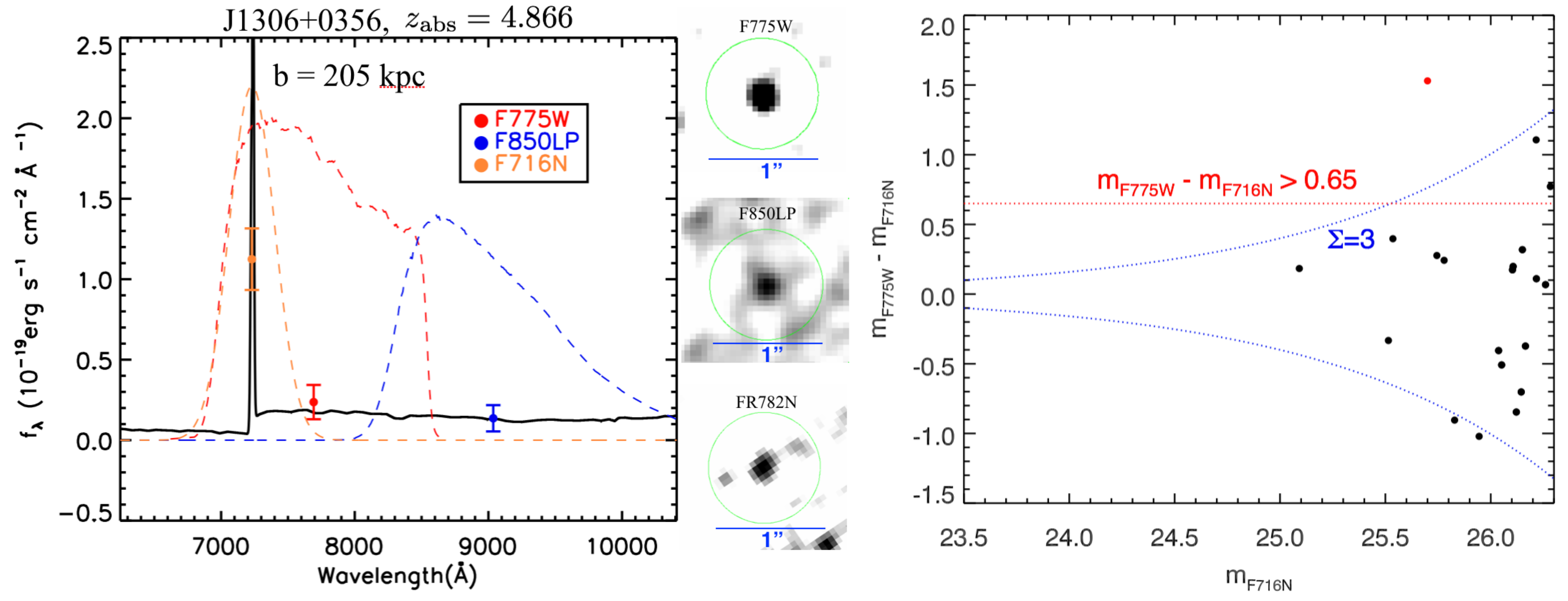}
\caption{Same format as the Figure~1, showing the photometry and the selection of LAE candidates  in the CIV absorber field at $z=4.866$. }
\end{figure}

\begin{sidewaystable} [!h]
\caption{Summary of  the CIV-associated Ly$\alpha$ emitter candidates} 
\label{table:PopIII_SFR}	
% is used to refer this table in the text
\centering 
\begin{tabular}{| c| c | c | c | c | c| c|c | c| c|c|c |c|c|} 
\hline\hline 
 RA & DEC  &  $z_{\rm{CIV}}$ &  Log[$\frac{\rm{N_{CIV}}}{\rm{cm}^2}$] & EW$_{\rm{CIV}}$ & Filter & NB &  F775W & F850LP &  $L_{\rm{Ly\alpha}}$ & SFR$_{\rm{Ly\alpha}}$ &  SFR$_{\rm{UV}}$ & b  & EW$_{\rm{Ly\alpha}}$  \\ 
 \hline
        &     &              &                     &               (\AA)  &     &     & & &     ($10^{42}$  erg s$^{-1}$) & (M$_\odot$ yr$^{-1}$) & (M$_\odot$ yr$^{-1}$)  & (kpc) & (\AA) \\
\hline
10:30:26.746  & +5:24:59.76 &  5.744 & $14.00\pm0.04$ &1.3  & FR853 & $25.90\pm0.21$ & $2\sigma\ge$28.0 & $27.10\pm0.13$ & 1.75$\pm0.44$ & 2.0 & 4.1 & 42  & 44 \AA \\  
\hline
   &    & 5.517 & $14.02\pm0.04$& 0.6                                      & FR782 &  $2\sigma \ge 26.8$ & $2\sigma\ge27.9$ &$2\sigma\ge28.0$  & $3\sigma\le 1.3$ &  $3\sigma\le 1.2$ &  $3\sigma\le 1.5 $  & $\ge125$ &   \\
\hline
10:30:27.486  & +05:25:20.15 & 4.948 & $13.76\pm0.11 $& 0.5 & FR716 & $25.24\pm0.12$ & $26.21\pm0.06$ & $26.04\pm0.06$ & $3.27\pm0.56$  & 3.0& 8.7 & 163  &  28 \AA \\
\hline
13:06:06.161 & +03:56:32.94 & 4.866 & $14.80\pm0.01$ & 2.6 & FR716 & $25.70\pm0.17$ &  $27.23\pm0.12$ &  $27.56\pm0.19$ & $2.72\pm0.55 $ & 2.5& 2.1 & 205 &  76 \AA \\
\hline
\end{tabular}
\end{sidewaystable}

\end{document}